\documentclass[useAMS,usenatbib]{mn2e}
\usepackage{graphicx}
\usepackage{color}
\usepackage{epsfig}
\usepackage{amsmath,amssymb}
\usepackage{natbib}
\usepackage{comment}

\def\diff{\mathrm{d}}
\def\ddt#1#2{\frac{\diff #1}{\diff #2}}
\def\pddt#1#2{\frac{\partial #1}{\partial #2}}

\def\simge{
    \mathrel{\rlap{\raise 0.511ex
        \hbox{$>$}}{\lower 0.511ex \hbox{$\sim$}}}}
\def\simle{
    \mathrel{\rlap{\raise 0.511ex
        \hbox{$<$}}{\lower 0.511ex \hbox{$\sim$}}}}

\newcommand{\secref}[1]{Section~\ref{#1}}

\newcommand{\msun}{M_{\odot}}

\newcommand{\ledd}{L_{\rm Edd}}
\newcommand{\fedd}{f_{\rm Edd}}
\newcommand{\mdotedd}{\dot{M}_{\rm Edd}}
\newcommand{\hii}{H{\sc ii}}
\newcommand{\yr}{{\rm yr} }

\newcommand{\dd}{{\rm d}}
\newcommand{\cc}{{\rm cm}^{-3}}
\newcommand{\K}{{\rm K}}

\defcitealias{Inayoshi2015a}{IHO16}

\title[Hyper-Eddington accretion onto luminous BHs]
{Hyper-Eddington mass accretion onto a black hole with super-Eddington luminosity}

\author[Yuya Sakurai, Kohei Inayoshi, Zolt\'an Haiman]
{Yuya Sakurai$^{1}$\thanks{sakurai@utap.phys.s.u-tokyo.ac.jp}, 
Kohei Inayoshi$^{2}$, Zolt\'an Haiman$^{2}$
\\
$^{1}$Department of Physics, School of Science, University of Tokyo, 
Bunkyo, Tokyo 113-0033, Japan\\
$^{2}$Department of Astronomy, Columbia University, 550 W. 120th Street, New York, NY 10027, USA}
\begin{document}

\date{Draft version \today}

\maketitle

\label{firstpage}

\voffset=-0.4in

\begin{abstract}
We perform one-dimensional radiation hydrodynamical simulations to
solve accretion flows onto massive black holes
(BHs) with a very high rate.  Assuming that photon trapping limits the
luminosity emerging from the central region to $L\la\ledd$,
Inayoshi, Haiman \& Ostriker (2016) have shown that an accretion
flow settles to a ``hyper-Eddington" solution, with a steady and
isothermal ($T\simeq 8000$~K) Bondi profile reaching $\ga 5000$ times
the Eddington accretion rate $\dot{M}_{\rm Edd}\equiv L_{\rm Edd}/c^2$.
Here we address the possibility that gas accreting with
finite angular momentum forms a bright nuclear accretion disc, with a
luminosity exceeding the Eddington limit ($1\la L/\ledd\la 100$). Combining our simulations with an 
analytic model, we find that a transition to steady hyper-Eddington accretion still
occurs, as long as the luminosity remains below
$L/\ledd\lesssim 35~(M_{\rm BH}/10^4~\msun)^{3/2}(n_\infty/10^5~{\rm 
cm^{-3}})(T_\infty/10^4~{\rm K})^{-3/2}(r_{\star}/10^{14}~{\rm cm})^{-1/2}$, 
where $n_\infty$ and $T_\infty$ are the density and
temperature of the ambient gas, and $r_\star$ is the radius of the
photosphere, at which radiation emerges. If the luminosity
exceeds this value, accretion becomes episodic. Our results
can be accurately recovered in a toy model of an optically thick
spherical shell, driven by radiation force into a
collapsing medium. When the central source is dimmer than the above
critical value, the expansion of the shell is halted and reversed by
ram pressure of the collapsing medium, and by shell's weight. Our results imply that rapid, unimpeded
hyper-Eddington accretion is possible even if the luminosity of the
central source far exceeds the Eddington limit, and can be either steady
or strongly episodic.
\end{abstract}


\section{Introduction}
\label{sec:introduction}

The existence of bright high-redshift ($z\ga 6$) quasars, powered by
supermassive black holes (SMBHs) with $\ga 10^{8-9}~\msun$, poses
questions about the formation and evolution of these SMBHs
\citep{Fan2003,Fan2006,Willott2010,Mortlock2011,Venemans2013,Wu2015}.

Several possible scenarios have been suggested for the origin of the
SMBHs \citep[][references therein]{Volonteri2010,Haiman2013,Johnson2016}.  One is remnant BHs
of massive Population III (Pop III) stars with $\sim 100~\msun$
\citep{Madau_Rees2001,Haiman_Loeb2001,Schneider2002,Islam2003,Volonteri2003,Tanaka_Haiman2009}.
Second, the so-called direct collapse model
(\citealt{Loeb_Rasio1994,Oh_Haiman2002,Bromm2003,Begelman2006,Natarajan2006,
  Shang2010a,Schleicher2013a,Regan2014,Inayoshi2014a,Visbal+2014,Alexander2014,
  Pacucci2015,Latif2015b}; \citealt{Chon2016})
considers a more massive seed BH with $\sim10^5~\msun$, formed by the
collapse of a supermassive star \citep[e.g.,
][]{Begelman2010,Hosokawa2013, Sakurai2015b}.  
Two direct-collapse BH candidates have recently been identifed in the CANDELS/GOODS-S survey based on their very red expected spectra \citep{Pacucci2016}, and may be detected in the future through stellar tidal disruption events \citep{Kashiyama2016}.
Thirdly, runaway collisions in star clusters produce massive stars which would be seeds
for the SMBHs (e.g., \citealt{Portegies_Zwart2004, Omukai2008,Devecchi_Volonteri2009,Katz2015,Yajima2016}; 
\citealt{Stone2016}).

How do seeds grow to SMBHs within the age of the universe at $z\ga6$?
In any of the above seed formation models, subsequent BH growth 
needs to be still rapid \citep{Tanaka2014}. 
When the BH is fed by sufficiently strong gas flows, and the emergent luminosity
increases, radiative feedback is likely to affect gas dynamics.
Radiation force is crucially important at the vicinity of the BH
horizon because the gas is highly opaque to electron scattering.  In
particular, if the luminosity approaches the Eddington luminosity,
$L_{\rm Edd}\equiv 4\pi cGM_{\rm BH}/\kappa_{\rm es}$, the radiation
force becomes comparable to the gravity of the accreting BH and
thus the accretion rate would be limited to the Eddington rate
$\dot{M}_{\rm Edd}\equiv L_{\rm Edd}/c^2$.  
Note that starting with a $100 (10^5)~\msun$ seed BH, it takes
  $\simeq 0.7 (0.4)$~Gyr to form a SMBH with $10^9~\msun$,
  assuming a continuous accretion at the Eddington rate with 10\%
  radiative efficiency, i.e., $\dot{M}=10~\dot{M}_{\rm Edd}$.
  A massive seed thus eases the requirement on the duty cycle by a factor for $z\ga 6$.
However, ``photon trapping" \citep{Katz1977,Begelman1978} would help BHs grow at a higher rate
than the Eddington rate.  Photon trapping occurs when radiation within
an optically thick flow is advected inwards by efficient electron
scattering faster than it can escape via radiation diffusion.  This
then limits the emergent luminosity, and, in spherical symmetry,
prevents it from exceeding the Eddington limit \citep{Begelman1979}.
The characteristic ``trapping radius'' is given by
\begin{equation}
R_{\rm tr}\equiv \frac{\kappa_{\rm es}\dot{M}}{4\pi c},
\label{eq:trapping}
\end{equation}
outside which radiation escapes and contributes to the emergent luminosity.
Thus, the maximum luminosity released by gravitational energy is estimated as 
$\simeq GM_{\rm BH}\dot{M}/R_{\rm tr}\sim L_{\rm Edd}$.
Numerical simulations have investigated rapid accretion with $\dot{M}\gg L_{\rm Edd}/c^2$
and found that such high accretion rates are possible in a disk-like configuration, with
radiation escaping vertically 
(e.g. \citealt{Ohsuga2005,Sadowski2014,Jiang2014,Mckinney2014,Fragile2014}; \citealt{Sadowski2016}).
Semi-analytical models also suggest the possibility of such rapid growth of SMBHs in the early Universe 
(e.g. \citealt{Volonteri2005,Madau2014,Alexander2014}; \citealt{Volonteri_Silk_Bubus2015}).
Analytical arguments also support rapid growth via gas accretion at a super-Eddington rate
\citep[e.g.][]{Pacucci_VF2015}.
A recent paper \citep{Pezzulli2016} 
shows that a high accretion rate of $\ga 10^3~L_{\rm Edd}/c^2$ is sustained at $z>10$,
even including a BH feedback model.
On the other hand, radiation heating potentially suppresses gas inflows from larger scales where gas is not bounded by the BH gravity. 
The radiation heating effect from a BH makes the accretion behavior episodic 
and the averaged rate value is limited to $\la 10\dot{M}_{\rm Edd}$
\citep[e.g.][]{Ciotti_2001,Milosavljevic2009b, Park_Ricotti2011,Park_Ricotti2012,Park2016}.
To address this issue, we need to consider {\it a self-consistent solution of the accretion flow
from larger scales where the gas accretion begins to small scales 
where photon trapping reduces the emergent luminosity}.

Recently, \citet{Inayoshi2015a} (hereafter IHO16) has found a steady self-consistent spherically symmetric solution of 
gas accreting at a rate of $\ga 5000~L_{\rm Edd}/c^2$ (hereafter ``hyper-Eddington accretion") when
\begin{equation}
\left(\dfrac{n_{\infty}}{10^5~\cc} \right)\ga \left(\dfrac{M_{\rm BH}}{10^4~\msun}\right)^{-1}
\left(\dfrac{T_{\infty}}{10^4~\K}\right)^{3/2}, 
\label{eq:condition}
\end{equation}
where $n_{\infty}$ and $T_{\infty}$ are the density and temperature of
the ambient gas.  They have shown that this condition corresponds to
the HII region, generated by the ionizing radiation emerging from the
photosphere, being smaller than the Bondi radius.  For the
hyper-Eddington case, the solution then consists of a
radiation-dominated core, where photon trapping due to electron
scattering is important, and an accreting envelope which follows an
isothermal Bondi profile with $T\simeq 8000~\K$.  If photon trapping
suppresses the luminosity, emerging from the photosphere, to below
$\sim L_{\rm Edd}$, radiation from the central region does not 
stop the gas accretion from larger scales. In fact, the size of the HII
region remains much smaller than the Bondi radius, which results in a
high inflow rate, unimpeded by radiation feedback.

However, \citetalias{Inayoshi2015a} assumed that photon trapping is
effective, and that the luminosity at the photosphere is limited to
$L_{\rm Edd}$.  This assumption is not valid when the accretion flow
with nonzero angular momentum produces a compact nuclear disk,
potentially launching outflows or jets into polar regions \citep[e.g.,
][]{Ohsuga2005,Sadowski2014,Mckinney2014}.
Photon trapping in the polar regions has also been found less
efficient than in spherically symmetric flows, with magnetic buoyancy
facilitating the vertical escape of radiation \citep{Jiang2014}.  As a
result, the radiation luminosity in directions away from the disk
place can significantly exceed the Eddington luminosity.

In this paper, we investigate the impact of a high-luminosity central
source, with $L >\ledd$, on the spherical accretion flow at larger
radii, by performing one-dimensional radiation hydrodynamical
simulations.  We show that the transition to steady hyper-Eddington
accretion still occurs, as long as the radiation luminosity from the
central bright source is as small as $L \la 10~\ledd$.  This high-rate
steady flow is maintained because the ram pressure of the infalling
gas dominates the radiation force caused by the central nuclear
disk.  We also use a toy model of a momentum-driven shell embedded in
a collapsing gas cloud, to demonstrate that the effect of the ram
pressure significantly suppresses radiation feedback, in good
agreement with our simulations.

This paper is organized as follows. In \secref{sec:method}, 
we describe the setup of our simulations and the numerical methodology.
We show the results of our simulations in \secref{sec:results} and give analytical arguments 
to explain the results in \secref{sec:analytical}.
In \secref{sec:summary}, we discuss our results and summarize our conclusions.

\begin{figure}
\centering
 \includegraphics[ width=7.5cm]{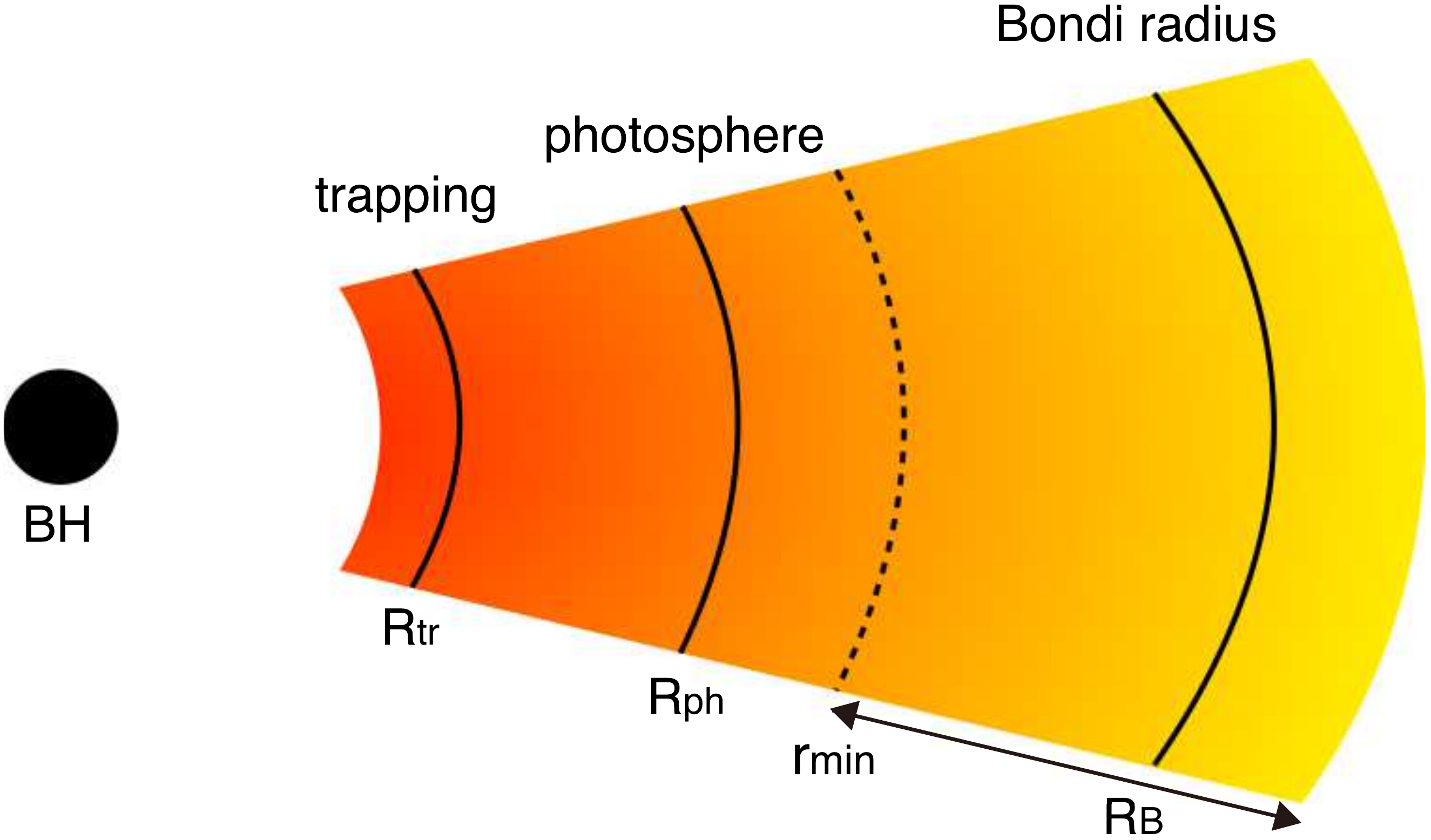}
 \caption{Schematic illustration of a spherically symmetric accretion flow
   onto a massive BH at a hyper-Eddington rate.  The locations of the
   trapping radius $R_{\rm tr}$, the photosphere $R_{\rm ph}$ and the
   Bondi radius $R_{\rm B}$ are shown.  Our simulations solve for the structure of the
   accretion flow between  $r_{\rm min}(\simeq
   10^{-3}~R_{\rm B})\lesssim r \lesssim 10~R_{\rm B}$.  }
 \label{fig:structure}
\end{figure}

\section{Simulation method}
\label{sec:method}
\subsection{Setup of the simulations}
\label{sec:suc}

We solve structures of spherical accretion flows onto a BH with 
a mass of $M_{\rm BH}$ by performing one-dimensional hydrodynamical 
simulations which include radiative processes.
Fig.~\ref{fig:structure} shows important physical scales of the gas structure.
The Bondi radius is estimated analytically by
\begin{equation}
R_{\rm B}=1.98\times10^{18}~M_{\rm BH,4}~T_{\infty,4}^{-1}~~{\rm cm},
\end{equation}
within which the BH gravity exceeds the gas pressure and accretion begins.
A photosphere forms at $R_{\rm ph}\sim 10^{14-15}$ cm (\citetalias{Inayoshi2015a})
and the trapping radius (equation \ref{eq:trapping}) is 
\begin{equation}
R_{\rm tr} = 1.48\times 10^{12}~M_{\rm BH,4}~\dot{m}_3~~{\rm cm},
\end{equation}
where $M_{\rm BH,4}\equiv M_{\rm BH}/(10^4~\msun)$, $T_{\infty,4}\equiv T_\infty/(10^4~{\rm K})$ 
and $\dot{m}_3\equiv (\dot{M}/\dot{M}_{\rm Edd})/10^3$.
It is desirable to resolve all these radii in the simulations in order to determine the structure of the flow self-consistently.
However, this is computationally prohibitive, because 
both $R_{\rm tr}$ and $R_{\rm ph}$ are smaller than the Bondi radius by $4-5$ orders of magnitude (see \citetalias{Inayoshi2015a}).
In the following simulations, therefore we focus on the region between
$10^{-3}~R_{\rm B}\lesssim r \lesssim10~R_{\rm B}$, and
investigate whether hyper-Eddington accretion is realized without being impeded by radiation feedback.
As a result, our simulation domain does not extend down to $R_{\rm tr}$ and $R_{\rm ph}$.
Instead, we set the emergent luminosity from the inner region by hand, using several different models of 
the radiation efficiency, including allowing for super-Eddington luminosities.
Note that in our paper, we simply assume that the disc is small enough to be embedded well inside the inner-most radius of the simulation box and consider emerging radiation with $L>\ledd$ from it.

\subsection{Basic equations and numerical schemes}
\label{sec:basic}

We use the hydrodynamical simulation code ZEUS \citep{Stone_Norman1992} to follow gas dynamics around the BH.
For the spherically symmetric case, the continuity equation is given by
\begin{equation}
\pddt{\rho}{t}+\frac{1}{r^2}\pddt{}{r}(r^2 \rho v)=0 \label{eq:continuity},
\end{equation}
and the equation of motion is given by
\begin{equation}
\rho\left(\pddt{v}{t}+v\pddt{v}{r}\right)=-\pddt{p}{r}-\rho\pddt{\Phi}{r}+f_{\rm rad} \label{eq:motion},
\end{equation}
where $\rho$ is the gas density, $v$ is the velocity of the flow, $p$ is the gas pressure, $\Phi$ is the gravitational potential of the BH, 
and $f_{\rm rad}$ is the radiation force per volume.
We assume the gas pressure is given by the equation of state $p=(\gamma-1)\rho e$, where $\gamma=5/3$ and 
$e$ is the specific energy density.
For completeness, we adopt the general relativistic correction for the gravitational potential, 
$\Phi=-GM_{\rm BH}/(r-R_{\rm Sch})$ \citep{Paczynsky1980a}, although in practice these corrections are negligible in our simulation domain.
The energy equation is given by
\begin{equation}
\rho\left(\pddt{e}{t}+v\pddt{e}{r}\right)=
-p\frac{1}{r^2}\pddt{}{r}(r^2v)-\Lambda+\Gamma \label{eq:energy},
\end{equation}
where $\Lambda$ is the cooling rate and $\Gamma$ is the heating rate.
The cooling rate $\Lambda$ includes the effect of the collisional excitation of H, He, He$^+$ atoms and 
by H free-free emission \citep{Glover2007}:
\begin{equation}
\Lambda=\Lambda_{\rm H}+\Lambda_{\rm He}+\Lambda_{\rm He^+}+\Lambda_{\rm ff}.
\end{equation}
The energy equation is solved by an implicit method, in order to stabilize the calculation.

We consider a chemical reaction network composed of the six primordial species of 
H, H$^+$, He, He$^+$, He$^{++}$ and e$^-$. 
The number abundance of He nuclei relative to H nuclei is set to $8.33\times10^{-2}$.
We consider the chemical reactions of photoionization, collisional ionization and radiative recombination of H, He, and He$^+$.
We adopt the on-the-spot approximation, i.e. recombination photons are quickly absorbed as ionizing photons and 
the recombinations to the ground state are ignored,
and we use the case B recombination coefficient.
The chemical reactions are solved for the six species with a semi-implicit formulation \citep{Anninos1997}.
The electron fraction is derived from charge conservation.

The time step of the simulation is set to the minimum value among 
the Courant time, the cooling time and the chemical time, following 
\citet{Whalen2006}.
We set the Courant number to be less than $0.5$. 
The cooling time and chemical time are set to the minimum value of 
\begin{equation}
t_{\rm cool}\equiv 0.1~\dfrac{\rho e}{| \Lambda-\Gamma |},
\end{equation}
\begin{equation}
t_{\rm chem}\equiv 0.01~\dfrac{x_{\rm e}+0.001~x_{\rm H}}{\dot{x}_{\rm e}},
\end{equation}
on the grid, where $x_{\rm e}$ and $x_{\rm H}$ are the electron and neutral hydrogen fraction, respectively.

We solve steady and spherically symmetric radiation transfer equations to calculate 
the radiation force, the heating rates and photoionization rates. 
The steady assumption is valid because the cloud crossing time of photons ($\tau r/c$) is much shorter than 
the simulation time step.
The transfer equation is given by
\begin{equation}
\frac{1}{r^2}\ddt{}{r}(r^2F_\nu)=4\pi \eta_\nu-\rho\kappa_\nu cE_\nu, 
\end{equation}
where $F_\nu$ is the radiation flux, $\eta_\nu$ is the emissivity, $\kappa_\nu$ is the opacity and $E_\nu$ is the radiation energy density.
The gas is optically thin against photons inside the ionized region and thus in those regions we approximate $F_\nu\approx cE_\nu$.

The photoionization rates $k_i$ and the heating rates $\Gamma_i$ ($i=$ H, He, He$^+$) are 
calculated with a photon-conserving scheme \citep{Whalen2006} 
\begin{align}
k_i&=\int_{\nu_i} \frac{4\pi \hat{J}_\nu}{h\nu}\sigma_i\dd\nu, \\
\Gamma_i&=n_i\int_{\nu_i} \frac{4\pi \hat{J}_\nu}{h\nu}\sigma_iE_{{\rm heat},i}\dd\nu,
\end{align}
where $\hat{J}_\nu$ is the mean intensity calculated to conserve the number of photons at each grid, 
$\sigma_i$ is the cross section for bound-free absorption of ionizing photons, 
$\nu_i$ is the ionization energy, and $E_{{\rm heat},i}=h\nu-h\nu_i$ is the excess energy of the photo-electron available for heating.
The radiation force due to electron scattering and bound-free transitions is calculated by 
\begin{equation}
f_{{\rm rad},i}=\frac{n_{\rm e}}{c}\int\sigma_{\rm es}F_\nu\dd\nu
+\frac{\Gamma}{c}
\end{equation}
where $\sigma_{\rm es}$ is the cross section for Thomson scattering.

\begin{figure}
\begin{center}
\includegraphics[width=75mm]
{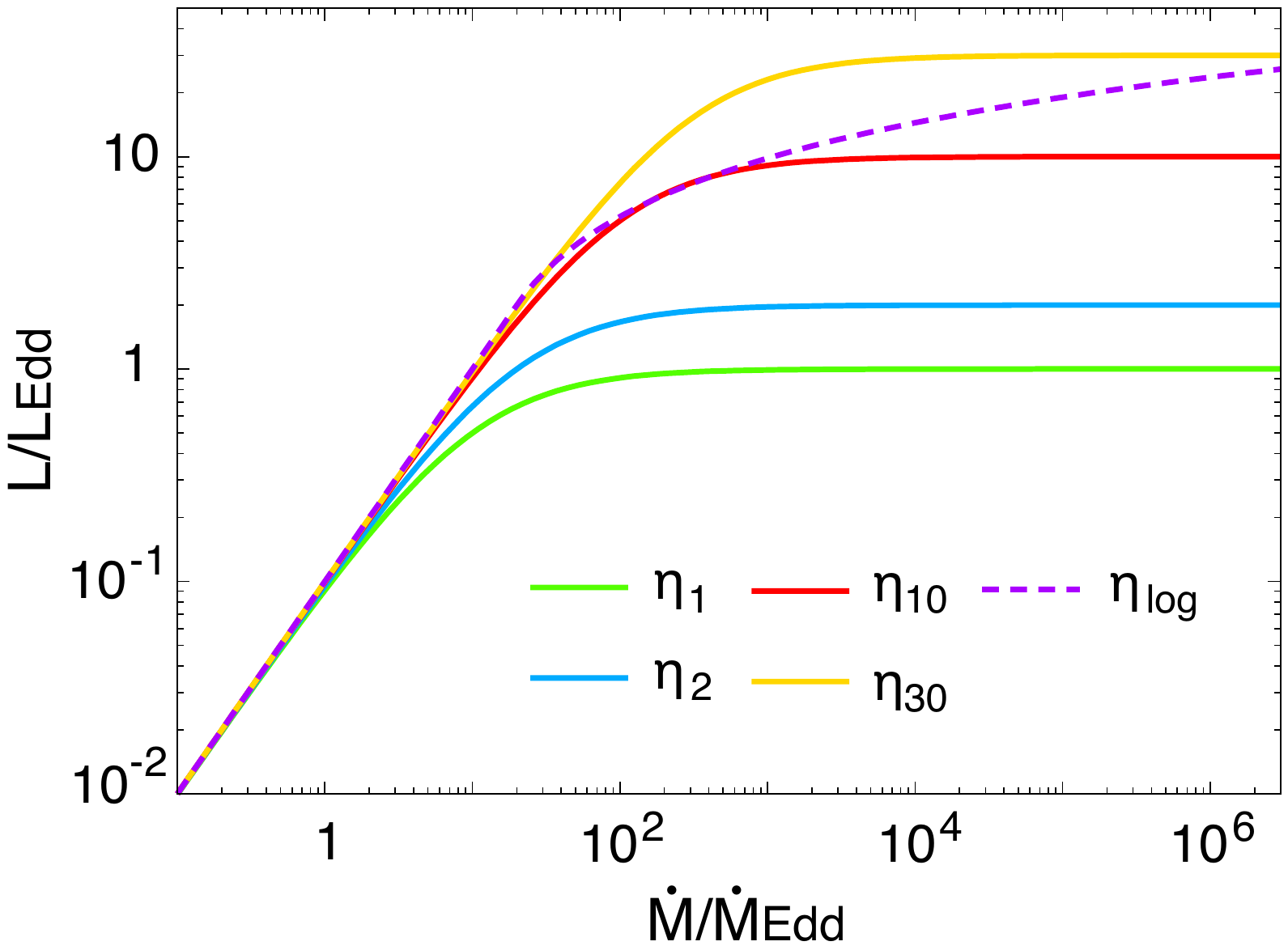}
\caption{ The bolometric luminosity assumed to be produced in the
  central regions, and enter our simulation domain at $r_{\rm min}$,
  as a function of the accretion rate $\dot{M}/\dot{M}_{\rm
    Edd}$. Different curves correspond to the different adopted models
  of the radiative efficiency, $\eta_{\fedd}$ with $1\leq \fedd \leq
  30$ (solid curves) and $\eta_{\rm log}$ (dashed curves).}
\label{fig:L_mdot}
\end{center}
\end{figure}

The radiation flux entering the simulation domain at its inner
boundary is specified by hand as follows.  The radiation spectrum is
assumed to be a single power-law of
\begin{align}
F_{\nu,{\rm in}}\propto \left(\frac{\nu}{\nu_{\rm min}}\right)^{-\alpha}~~(\nu_{\rm min}\le\nu\le\nu_{\rm max}), \label{eq:power}
\end{align}
where $\nu_{\rm min}$ is the ionization threshold of neutral hydrogen, $h\nu_{\rm min}=13.6~{\rm eV}$, and 
$h\nu_{\rm max}\sim30~{\rm keV}$ 
is the maximum cutoff frequency.
The power-law index is set to $\alpha=1.5$ (see IHO16).
The normalization of the radiation flux is determined by $L=\eta\dot{M}c^2$, where
$\dot{M}$ is the mass flux through the innermost grid, $\eta$ is the radiative efficiency, and $L$ is the bolometric luminosity.
We assume a simple model of the efficiency which mimics the  effect of photon trapping for a high $\dot{m}(\gg 1)$ as
\begin{equation}
\eta_{\fedd}\equiv \frac{1}{10+\dot{m}/\fedd},
\label{eq:eta_model}
\end{equation}
where $\fedd\equiv L_{\rm max}/\ledd$ and $L_{\rm max}$ is the maximum luminosity for $\dot{m}\rightarrow \infty$ (see Fig.~\ref{fig:L_mdot}). 
In this model, the efficiency becomes a constant ($\eta \simeq 0.1$) for low $\dot{m}$, while $\eta \rightarrow f_{\rm Edd}/\dot{m}$ for high $\dot{m}$.
  We do not consider an advection-dominated accretion flow (ADAF), because the accretion rates in
  our simulations do not drop below the critical value, $\dot{m}\approx 10^{-3}$ \citep{Ichimaru1977,Narayan1994},
  at which a transition to ADAF would be expected (see Fig.~\ref{fig:mdot1}).
Note that \citetalias{Inayoshi2015a} considered only the case with $f_{\rm Edd}=1$, where the luminosity never exceeds the Eddington luminosity.
However, we here relax this assumption and allow super-Eddington luminosities.
In addition to this model, we also consider a model of the efficiency which asymptotically approaches a logarithmic form at high $\dot{m}$ 
\citep[][]{Watarai2000},
\begin{align}
\eta_{\rm log}=
\begin{cases}
0.1 & (\dot{m}<20) \\
\displaystyle\frac{2}{\dot{m}}\left[1+\ln\left(\frac{\dot{m}}{20}\right)\right] & (\dot{m}>20).
\end{cases}
\label{eq:eta_log}
\end{align}
This prescription is motivated by the simulations with $\dot{m}\gg 1$, mentioned above, which find
that photon trapping do not fully suppress the luminosity  emerging from the central region \citep[e.g.,][]{Jiang2014,Sadowski2014}.

We set spherical coordinates with a logarithmically-spaced grid in the radial direction as follows. 
The positions of the inner and outer boundary are set to $r_{\rm min}$ and $r_{\rm max}$.
The $i$-th grid is given by $r_i=r_{\rm min}+\Delta r_0(\epsilon^{i-1}-1)/(\epsilon-1)$,
where $\Delta r_0$ is the size of the inner-most grid and $\epsilon~(=\Delta r_{\rm i+1}/\Delta r_i)$ is the ratio between consecutive grids.
For a given number of the grid-cells $N$, $\Delta r_0=(r_{\rm max}-r_{\rm min})(\epsilon-1)/(\epsilon^N-1)$.
Throughout this paper, we set $N=700$, $\epsilon=1.01$, 
$r_{\rm min}\sim 10^{-3}~R_{\rm B}$ and $r_{\rm max}=5000~r_{\rm min}$ 
so that dynamics of gas accretion from outside the Bondi radius are calculated with sufficient resolution.

We adopt the ``outflow'' boundary condition (BC) at the innermost grid \citep[e.g.][]{Stone_Norman1992}.
Under this BC, we set $v(r_{\rm min})=v(r_{\rm min}+\Delta r_0)$ to avoid spurious reflection 
of wave energy at the boundary.
However, when $L> \ledd$, this BC artificially underestimates the effect of the radiation force on the innermost shell.
This is because the velocity at $r_{\rm min}$, where the infalling gas should be significantly decelerated by radiation, is replaced by
the velocity at $r_{\rm min}+\Delta r_0$, where deceleration is inefficient because radiation is partially
absorbed by the gas at $r_{\rm min}$ before reaching $r_{\rm min}+\Delta r_0$.
To circumvent this underestimate, we adopt an alternative inner BC: $v(r_{\rm min}+\Delta r_0)=v(r_{\rm min})$ for $L> \ledd$.
When we use this BC, we choose the size of the innermost grid so that the gas element 
at $[r_{\rm min},~r_{\rm min}+\Delta r_0]$ is optically thick to electron scattering ($\tau_{\rm es}\ga 1$).
If we chose a small grid size of $\Delta r_0$ so that $\tau_{\rm es}\ll 1$, 
most of radiation should penetrate to the second grid and cause significant deceleration there.
This would be missed by the new BC of $v(r_{\rm min}+\Delta r_0)=v(r_{\rm min})$, which would then again underestimate
the effect of radiation force. 
We have checked that $\tau_{\rm es}\ga 1$ at $r=r_{\rm min}+\Delta r_0$ is ensured for $N=700$.

\section{Results}
\label{sec:results}
\begin{figure}
\begin{center}
\includegraphics[width=80mm]
{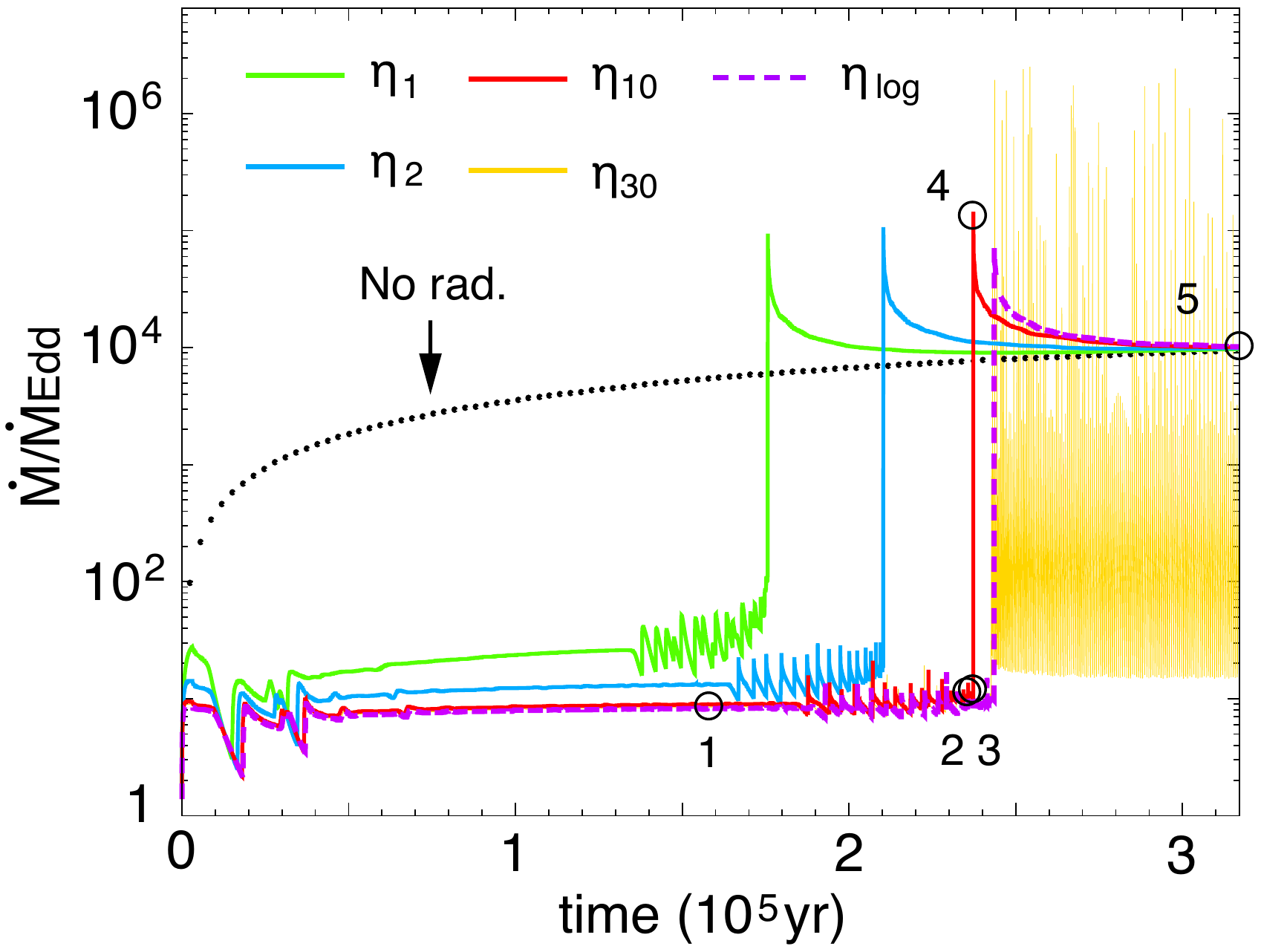}
\caption{
Time evolution of the accretion rates for the five different models of 
radiative efficiency shown in Fig.~\ref{fig:L_mdot}
(and for $M_{\rm BH}=2\times10^4~\msun$, $n_\infty=10^5~{\rm cm^{-3}}$ and 
$T_\infty=10^4~$K).
The dotted curve shows the time evolution of the rate in the absence of any radiation (settling to the Bondi rate).
Circles mark five different epochs, at which we show radial profiles 
for the model of $\eta_{10}$ in Fig. \ref{fig:hist1}.
}
 \label{fig:mdot1}
\end{center}
\end{figure}

Fig.~\ref{fig:mdot1} shows the time evolution of the accretion rate for several models of the radiative efficiency
$\eta_{f_{\rm Edd}}$ ($1\leq f_{\rm Edd}\leq 30$) (solid curves) and $\eta_{\rm log}$ (dashed curve).
We here set the BH mass to $M_{\rm BH}=2\times10^4~\msun$ and adopt an initially neutral uniform gas with
$n_\infty=10^5~{\rm cm^{-3}}$, $T_\infty=10^4~{\rm K}$ and $v=0$.
The dotted curve presents the evolution with radiation off at the inner boundary, 
approaching the Bondi rate,
\begin{align}
\frac{\dot{M}_{\rm B}}{\mdotedd}
=7.3\times10^3~M_{\rm BH,4}n_{\infty,5}T_{\infty,4}^{-3/2},
\label{eq:Bondi}
\end{align}
where we estimate the Bondi rate for the isothermal case, $\dot{M}_{\rm B}=e^{3/2}\pi\rho_\infty R_{\rm B}^2 c_\infty$.
With radiation on for $\fedd=1$, the accretion rate is much lower than the case without radiation.
The average rate is limited to $\sim 20~\mdotedd$ at $t<10^5$ yr, 
where the luminosity from the central region is $\sim 0.7~L_{\rm Edd}$.
At $t \ga 1.3\times 10^5$ yr, the accretion rate becomes episodic and increases promptly to a higher value, which we call the transition.
This result is consistent with that found in \citetalias{Inayoshi2015a} (see their Fig. 5).
After the transition, the accretion rate approaches the Bondi rate.
The hyper-Eddington accretion is realized because the \hii~region is always confined inside the Bondi radius,
i.e. $R_\text{\hii}\la R_{\rm B}$ (see below and \S\ref{sec:analytical}).

For moderately larger values of $1<f_{\rm Edd}\leq 10$, we find the same transition
to steady hyper-Eddington accretion as in $f_{\rm Edd}=1$.
Even in these cases, the luminosity before the transition is limited to $\sim L_{\rm Edd}$.
The transition time is delayed for higher $f_{\rm Edd}$ because radiation force is non-negligible,
and contributes an outward-directed force on the gas.
After the transition, the luminosity exceeds the Eddington luminosity.
However, the hyper-Eddington accretion is maintained since the ram pressure overcomes 
the radiation force at the innermost region. 
Note that the result in the model with $\eta_{\rm log}$ (equation \ref{eq:eta_log}) does not change qualitatively
because the luminosity after the transition ($L\la 20~L_{\rm Edd}$) is as small as in the cases 
with $1<f_{\rm Edd}\leq 10$.

For the highest value of $f_{\rm Edd}=30$, the transition to a hyper-Eddington phase occurs, 
but the behavior of the accretion rate is different from the other cases after the transition.
Namely, the accretion rate is unstable, and begins to oscillate
at the innermost grid.
In this case, radiation force with $L\simeq 30~L_{\rm Edd}$ from the central region 
prevents a steady accretion flow from being realized. 
However, the radiation force does not suppress the gas accretion in the quiescent phases.
As a result, the time-averaged accretion rate still matches the Bondi rate.
This implies that the central BH grows rapidly even for $\fedd \ga 30$ 
(see discussion in \S\ref{sec:summary}).
This critical luminosity to maintain steady hyper-Eddington accretion is determined by 
a comparison of the radiation force with the ram pressure and gravity of the infalling gas (see \secref{subsec:analytical1} below).

\begin{figure}
\begin{center}
\includegraphics[width=80mm]
{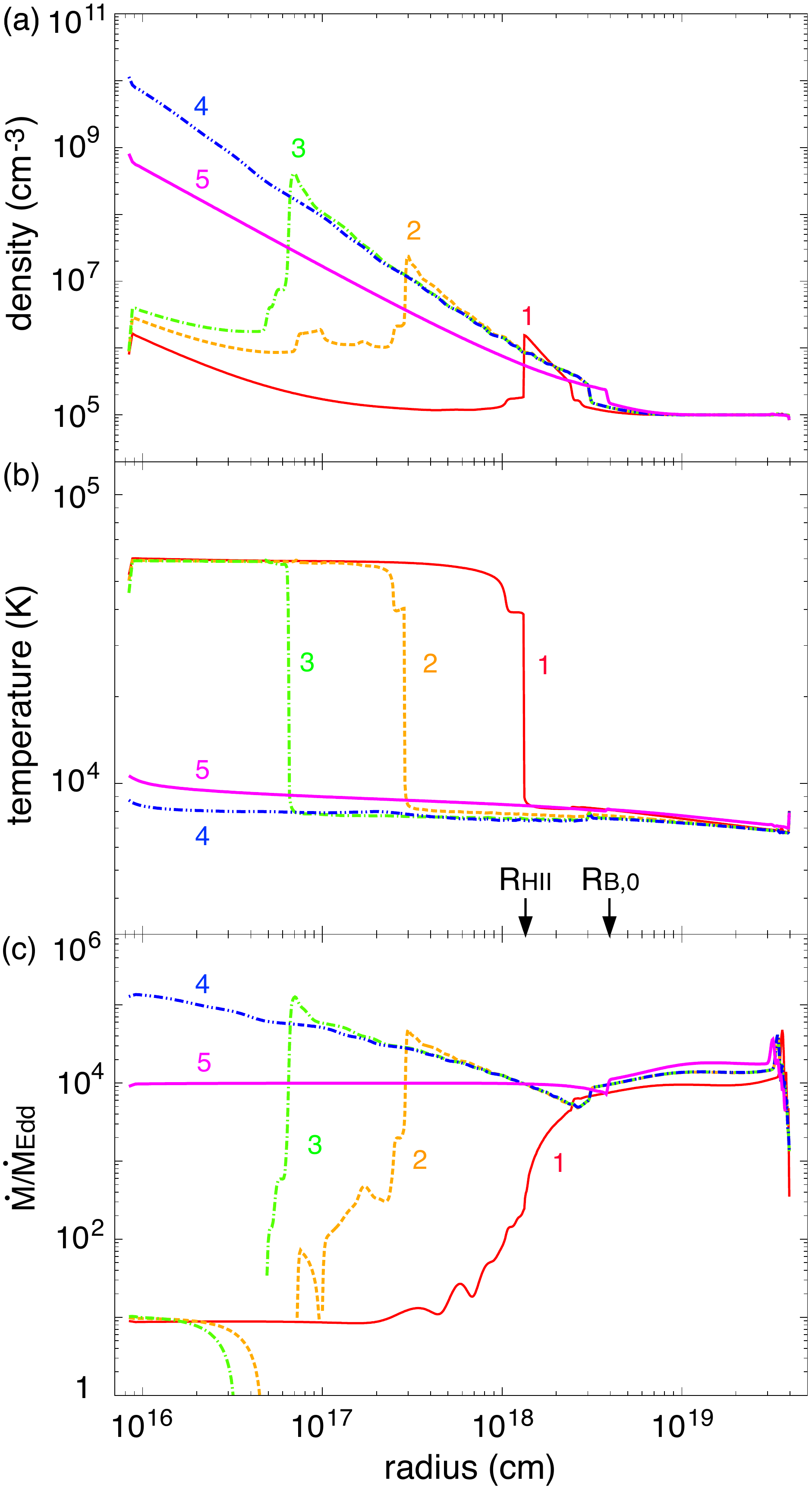}
\caption{
Radial profiles of the density, temperature and local accretion rate for the model of $\eta_{10}$ 
 (see equation \ref{eq:eta_model})
 at the five different epochs shown by circles in Fig.~\ref{fig:mdot1}.
 The five curves correspond to (i) $t=1.58\times10^5~\yr$ (red), (ii) $2.35\times10^5~\yr$ (orange), 
 (iii) $2.371\times10^5~\yr$ (green), (iv) $2.373\times10^5~\yr$ (blue) 
 and (v) $3.17\times10^5~\yr$ (magenta).
 In panel (b), the size of the \hii~region $R_\text{\hii}$ and the initial Bondi radius $R_{\rm B,0}$ are shown.
 In this case, the condition required for steady hyper-Eddington accretion, $R_\text{\hii}\la R_{\rm B,0}$,  is satisfied.
}
  \label{fig:hist1}
\end{center}
\end{figure}

Fig.~\ref{fig:hist1} shows radial profiles of the gas density, temperature, and local mass inflow rate $\dot{M}=4\pi r^2\rho |v|$
for the model with $\fedd=10$ at five different epochs corresponding to the filled circles in Fig.~\ref{fig:mdot1}.
As \citetalias{Inayoshi2015a} discussed, the condition required for hyper-Eddington accretion 
is given by equation (\ref{eq:condition}), which is equivalent to the condition that
the size of the \hii~region is smaller than the initial Bondi radius 
(see also \S\ref{subsec:analytical1}).
Fig.~\ref{fig:hist1}(b) shows that the condition is satisfied for this case ($R_\text{\hii}<R_{\rm B,0}$).
The physical explanation why this transition occurs are as follows.
When the accretion occurs, the \hii~region expands to the radius of $R_\text{\hii}$, 
within which radiation force and gas pressure suppress the gas accretion.
However, since $R_\text{\hii}<R_{\rm B,0}$, the accreting gas accumulates
in the region $R_\text{\hii} \la r \la R_{\rm B}$ (curve 1 in Fig.~\ref{fig:hist1}a).
Once this shell becomes sufficiently dense and massive, it begins to fall inward 
due to the gravitational force of the central BH (curves 2 and 3 in Fig.~\ref{fig:hist1}a).
Concurrently, the \hii~region shrinks and the accretion rate increases (curve 4 in Figs.~\ref{fig:hist1}b and c).
After the transition, the gas profile approaches a steady and isothermal Bondi profile
with $\rho \propto r^{-3/2}$ and $T\simeq 8000~\K$ (curve 5 in Fig.~\ref{fig:hist1}).

For the case with $\fedd=30$, radial profiles of the gas properties are almost identical to those for $\fedd=10$,
except inside a narrow central \hii~region.
In this case, the strong radiation force eventually blows the ionized gas outward.
However, once the luminosity decreases due to suppression of the gas accretion,
ram pressure caused by rapid accretion from outside the \hii~region
pushes the ejected gas inward again, resulting in episodic accretion.
This result also shows that as long as $R_\text{\hii}<R_{\rm B,0}$ is satisfied,
the time-averaged accretion rate should match the steady hyper-Eddington Bondi rate,
even for $\fedd=30$.

Finally, we ran a separate, simpler suite of simulations, as an academic exercise, 
to further clarify the effect of radiation force with a super-Eddington luminosity on the accretion flow.
In the above simulations, the luminosity is coupled with the accretion rate.
Instead, we here assume a steady and isothermal Bondi accretion flow as the initial condition, 
and turn on the central source with a {\em constant} luminosity, independent from the the
accretion rate.  This setup allows us to compare the results directly with a toy model, discussed in \S\ref{subsec:analytical2}
below.
Fig.~\ref{fig:mdot2} shows two cases for  $L=10~\ledd$ (red solid) and $30~\ledd$ (blue dashed).
For $L=10~\ledd$, the gas accretion does not change at all after the radiation turns on at $t=0$.
On the other hand, for $L=30~L_{\rm Edd}$, the radiation force is so strong that the gas inflow is decelerated and shuts off.
This behaviour differs from the previous case with $\fedd=30$ (see Fig.~\ref{fig:mdot1}), in which
the radiation force was set to depend on the accretion rate, and accretion was episodic, rather than being shut off.

\section{Analytic arguments}
\label{sec:analytical}

As we have shown in \S\ref{sec:results}, a transition to a steady
hyper-Eddington accretion occurs for $M_{\rm BH}=2\times 10^4~\msun$
and $n_\infty=10^5~\cc$, as long as the maximum luminosity from the
central region is $L/L_{\rm Edd}(=f_{\rm Edd})\la 20$.  Although a
steady accretion is replaced by strongly fluctuating, episodic
accretion for $\fedd \geq 30$, the time-averaged rate remains close to
the Bondi rate.  We discuss the physical interpretation of these
results with a toy model of an optically thick spherical
shell, driven by radiation force from a central source.

\subsection{Hyper-Eddington accretion conditions}
\label{subsec:analytical1}

As we explain in \S\ref{sec:results} (see also \citetalias{Inayoshi2015a}), 
the transition to hyper-Eddington accretion occurs when the Bondi radius $R_{\rm B}(\propto M_{\rm BH}T_\infty^{-1})$ is 
larger than the size of the \hii~region $R_\text{\hii}$. The latter is estimated by the balance between photoionization and radiative recombination as
\begin{equation}
R_\text{\hii}=\left(\frac{3Q_{\rm ion}}{4\pi n_\infty^2\alpha_{\rm B}}\right)^{1/3}, \label{eq:r_hii}
\end{equation}
where $Q_{\rm ion}$ is the mean number of ionizing photons emitted per unit time 
and $\alpha_{\rm B}$ is the case B recombination rate.
Since we consider the power-law spectrum with the index of $-1.5$, 
we obtain $Q_{\rm ion}=L/(3h\nu_{\rm min})$.
Before the transition occurs, the luminosity is limited to $\sim \ledd$
(see Figs.~\ref{fig:L_mdot} and \ref{fig:mdot1}).
Thus, since $R_\text{\hii} \propto L_{\rm Edd}^{1/3}n_\infty^{-2/3}\propto M_{\rm BH}^{1/3}n_\infty^{-2/3}$,
the transition condition of $R_{\rm B}>R_\text{\hii}$ is written as equation (\ref{eq:condition})
where we set the temperature within the \hii~region to $6\times 10^4~\K$.
Note that for the evaluation of $R_\text{\hii}$ we here assume the constant density profile with $n_\infty$
instead of the Bondi density profile, which is actually realized just before the transition. 
The resulting value of $R_\text{\hii}$ from Eq. \eqref{eq:r_hii} is larger by a factor of a few than the actual value 
because equation (\ref{eq:r_hii}) neglects the fact that the density profile has a steep slope 
($\rho \propto r^{-\beta}$; $0<\beta<3/2$).
Thus, our assumption of a constant density profile is rather conservative in terms of the conditions for hyper-Eddington accretion.

\begin{figure}
\begin{center}
\includegraphics[width=80mm]
{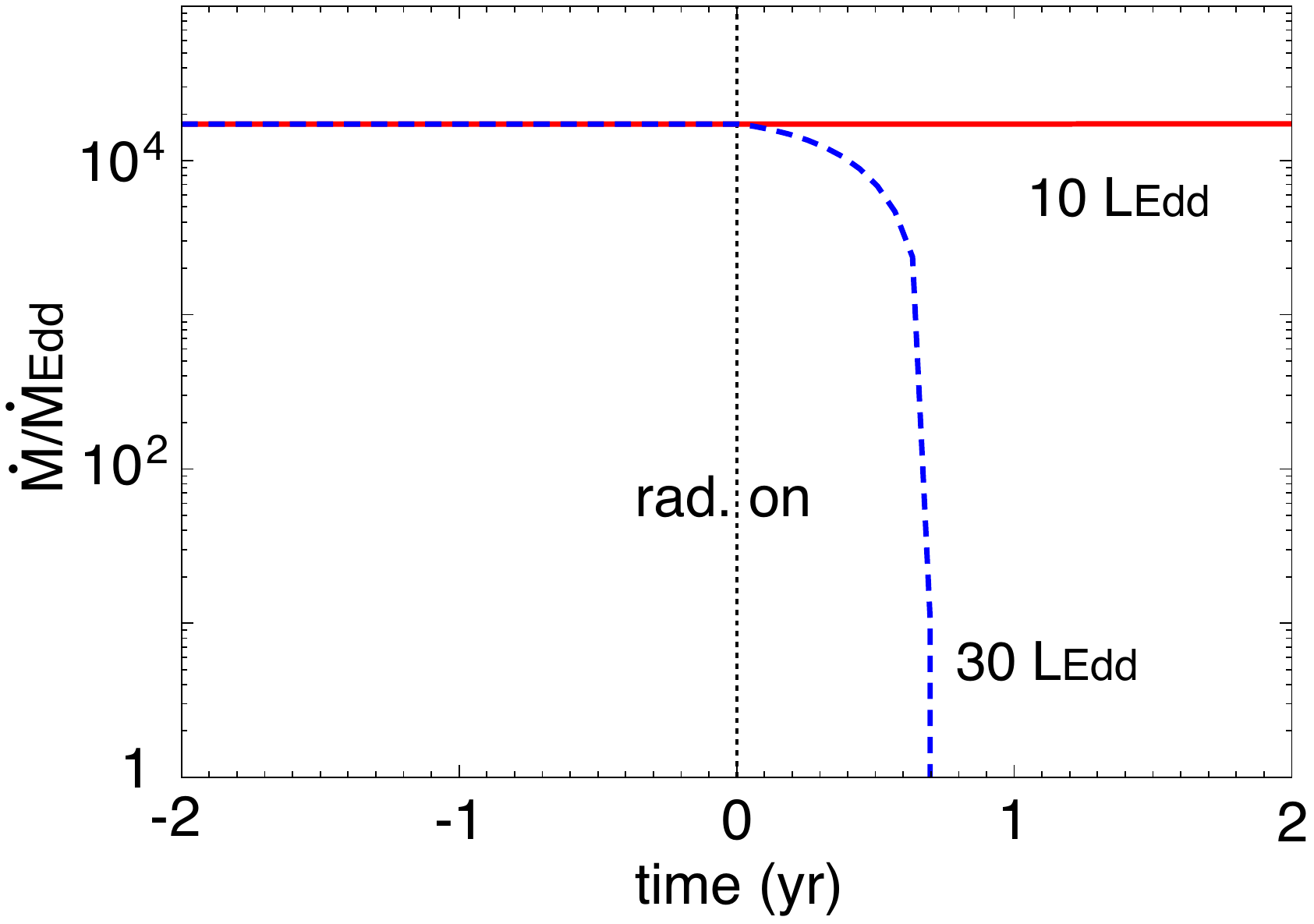}
\caption{
The effect of radiation force from a constant luminosity source on an initially steady and isothermal Bondi accretion flow.
At $t=0$, we turn on the radiation with $L=10~\ledd$ (red solid) or $L=30~\ledd$ (blue dashed).
In the former case, the accretion rate does not change from the Bondi rate, 
while in the latter case, the accretion is shut off by radiation force.}
  \label{fig:mdot2}
\end{center}
\end{figure}

After the transition, the radiation luminosity from the central BH in certain directions
would be brighter than $\sim L_{\rm Edd}$
\citep[e.g.][]{Ohsuga2005,Sadowski2014,Jiang2014}.
In the standard picture of outflows driven by radiation force with $L>L_{\rm Edd}$,
hyper-Eddington accretion seems unlikely to occur 
because radiation force due to electron scattering dominates the BH gravity.
However, in our case, all momentum of the radiation is essentially
absorbed by neutral hydrogen at the edge of the \hii~region within a
short mean-free path.  As a result, the radiation force,
exerted on the recombination shell near $R_\text{\hii}$ is in fact {\em
  larger} than that onto ionized gas by a factor of $1/\tau_{\rm e}$,
where $\tau_{\rm e}\sim n\sigma_{\rm T}R_\text{\hii}~(\lesssim1)$.  On the other
hand, the radiation has no impact outside the \hii~region, where rapid hyper-Eddington inflow can develop.
As a result, a large ram pressure is exerted inward at the boundary of the \hii~region, which 
can significantly exceed the gravity of the BH. Furthermore, the infalling gas can accumulate near
$R_\text{\hii}$ and increase the inward gravitational force.
Therefore, the usual calculation of the Eddington limit, which equates radiation force on electrons with the BH's gravity,
must be replaced in our case by a comparison between the (larger) radiation force on the neutral HI
and the (larger) inward ram pressure and gravity.

When the ram pressure dominates the radiation force even after the transition,
a steady hyper-Eddington accretion is maintained.
The stability condition is written as $\dot{M}_{\rm B}|v|>L/c$ at $r=r_\star$ 
where all radiation is absorbed.
Since the inflow velocity is set to $|v|=(2GM_{\rm BH}/r)^{1/2}$ at $r\geq r_\star$,
we obtain
\begin{equation}
\fedd =\frac{L}{\ledd}\la 11~M_{\rm BH,4}^{3/2}~n_{\infty,5}~T_{\infty,4}^{-3/2}~r_{\star,15}^{-1/2}, 
\label{eq:ram}
\end{equation}
where $r_{\star,15}=r_\star/(10^{15}~{\rm cm})$.  As a conservative
estimate, we set $r_\star=r_{\rm min}(=8\times10^{15}~{\rm cm})$.  For
the case with $M_{\rm BH,4}=2$, $n_{\infty,5}=1$ and $T_{\infty,4}=1$,
hyper-Eddington accretion remains stable as long as $\fedd\lesssim10$.
This estimate agrees with our simulation results shown in
\S\ref{sec:results}.  In practice, the radiation should emerge from the
photosphere located at a smaller radius, $R_{\rm ph}(<r_{\rm min})$.
Although our simulations do not resolve the photosphere, if we adopt
$R_{\rm ph}\simeq 10^{14}~{\rm cm}$ (shown in Fig. 11 of
\citetalias{Inayoshi2015a}), we find the critical luminosity of
$\fedd\lesssim100$.  We discuss this critical luminosity further, using a
simple toy model, in \S\ref{subsec:analytical2} below.

\begin{figure}
\begin{center}
\includegraphics[width=80mm]
{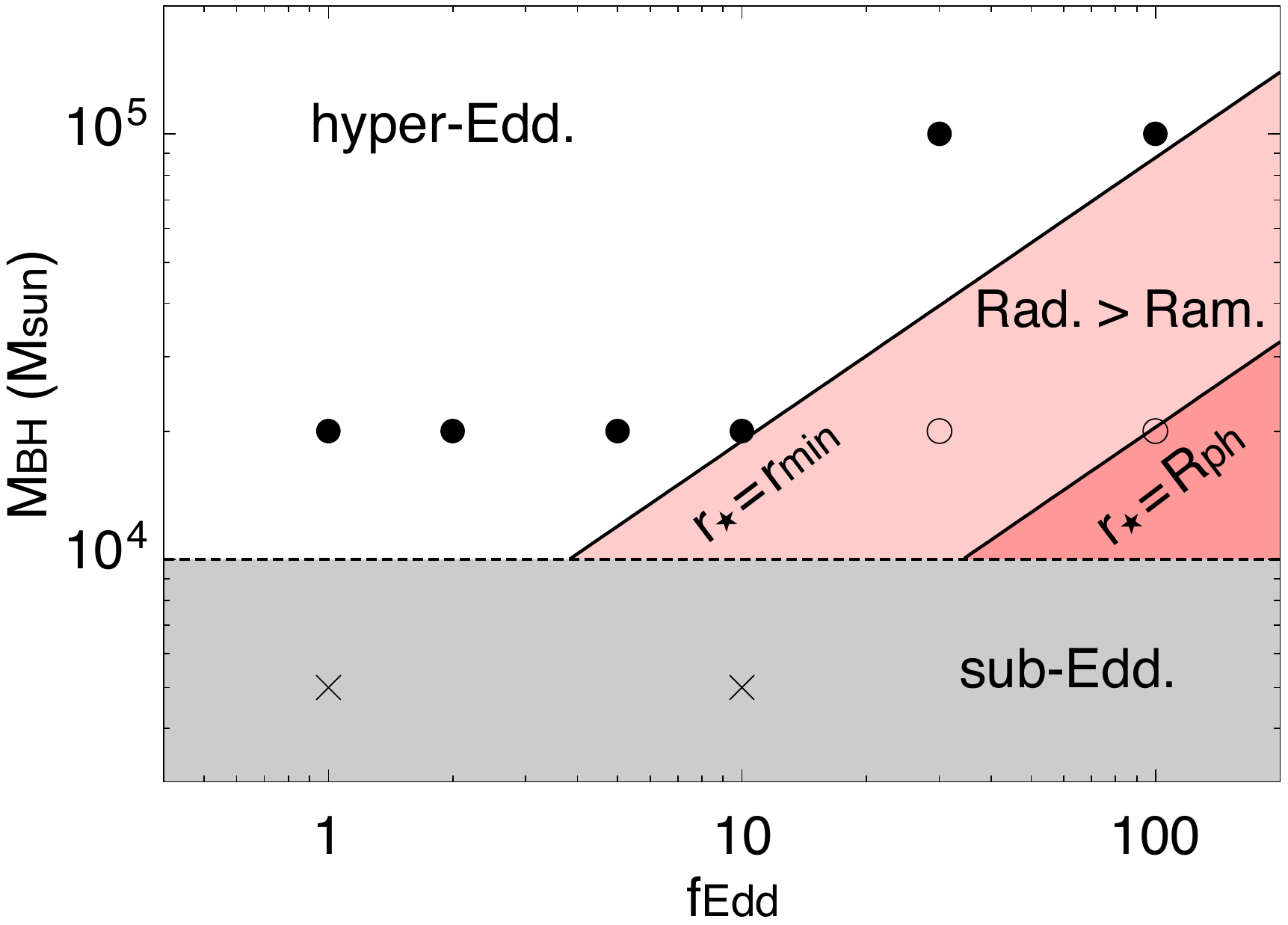}
\caption{Summary of three different accretion regimes, occurring for
  different values of $\fedd$ and $M_{\rm BH}$.  Dashed and solid
  lines show the conditions given by Eqs. (\ref{eq:condition}) and
  (\ref{eq:ram}) for $n_\infty=10^5~{\rm cm^{-3}}$,
  $T_\infty=10^4~{\rm K}$, respectively.  For the solid lines, we show
  two cases with the central radiation emerging from $r_\star=r_{\rm
    min}$ (the inner boundary of the simulations) or from $R_{\rm ph}$
  (the expected location of the photosphere).  Different symbols mark
  simulation runs in which steady hyper-Eddington accretion was
  realized (filled circles), accretion became strongly episodic due to
  radiation force (open circles), and the transition to
  hyper-Eddington flow was suppressed entirely because of radiation
  heating and ionization (crosses).}
  \label{fig:condition1}
\end{center}
\end{figure}

\begin{figure}
\begin{center}
\includegraphics[width=80mm]
{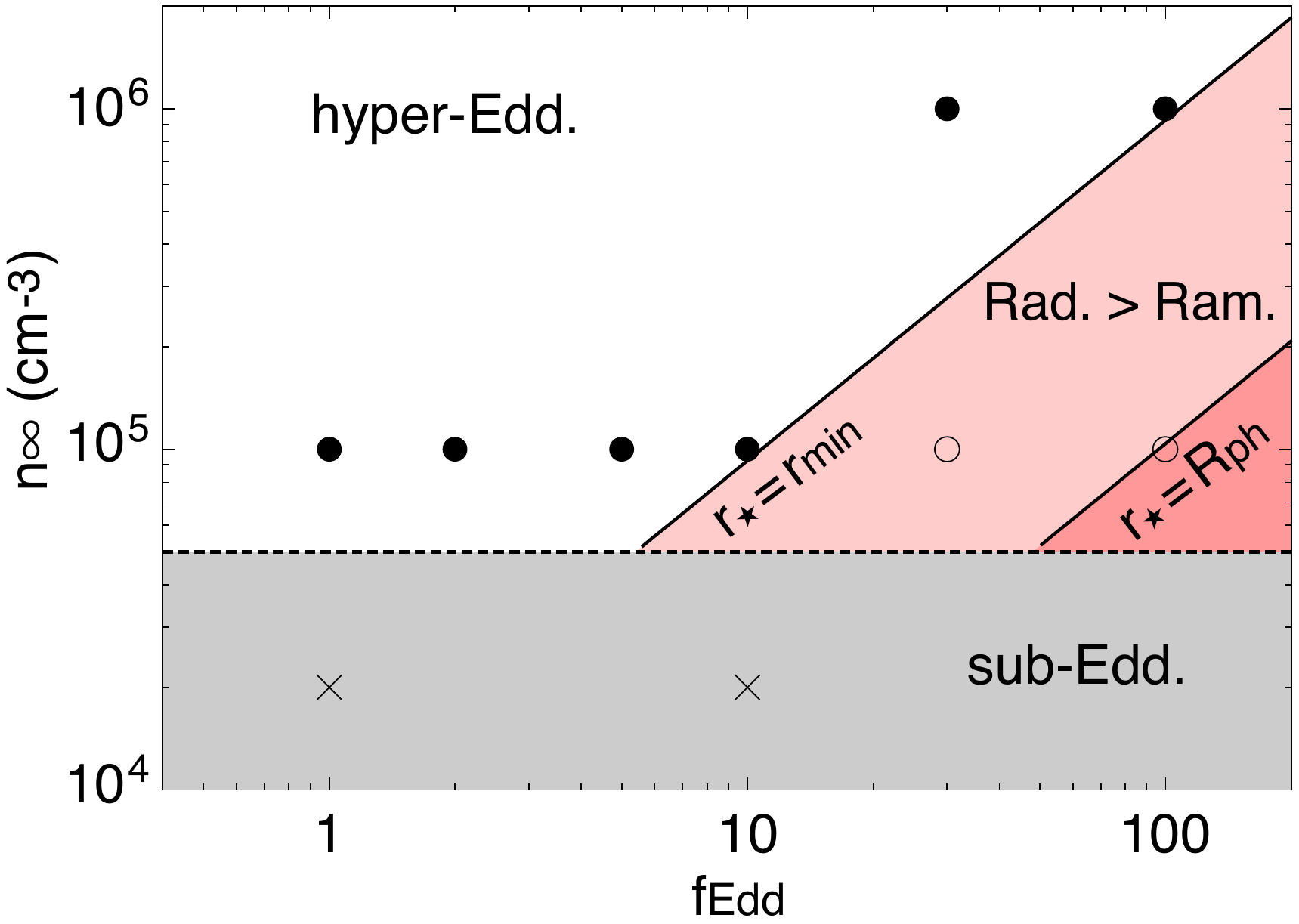}
\caption{
Same figure as Fig. \ref{fig:condition1} but for $\fedd$ and $n_\infty$. 
 We set $M_{\rm BH}=2\times10^4~\msun$ and $T_\infty=10^4~{\rm K}$.
}
  \label{fig:condition2}
\end{center}
\end{figure}

We summarize the necessary conditions for hyper-Eddington accretion 
in Figs.~\ref{fig:condition1} and \ref{fig:condition2}.
The conditions of equations (\ref{eq:ram}) (solid) and (\ref{eq:condition}) (dashed) are shown 
in the $\fedd-M_{\rm BH}$ and $\fedd-n_\infty$ planes, respectively.
For the solid lines, we have set either $r_\star=r_{\rm min}$ or $r_\star=R_{\rm ph}$.
Below the dashed lines, hyper-Eddington accretion is not realized
because of radiation heating and ionization (cross; sub-Edd.).
In the region between the solid and dashed lines,
the hyper-Eddington accretion could occur but a steady state is not achieved 
because of radiation force dominating ram pressure and gravity
(open circle; Rad. $>$ Ram.).
Only in the region above those lines, a steady hyper-Eddington accretion is allowed (filled circle; Hyper-Edd.).

\subsection{1D model for a momentum-driven shell}
\label{subsec:analytical2}

In order to understand the physics which allows hyper-Eddington
accretion, we consider a toy model of a geometrically thin, but
optically thick spherical shell around a point source, driven by
radiation force into a rapidly collapsing medium
\citep[e.g.][]{King2003,Kasliwal2005}.  The luminosity $L$ of the central
source is assumed constant, and the equation of motion of the shell is given by
\begin{equation}
\ddt{}{t}(M_{\rm sh}\dot{R}_{\rm sh})=\frac{L}{c}-\dot{M}(|v|+\dot{R}_{\rm sh})-\frac{GM_{\rm BH}M_{\rm sh}}{R_{\rm sh}^2},
\end{equation}
where $M_{\rm sh}$ is the mass of the shell, $R_{\rm sh}$ is the distance of the shell from the center, 
and $\dot{M}$ and $v$ are the accretion rate and velocity of the gas inflow just outside the shell.
The terms on the right hand side correspond to the outward force exerted on the shell by radiation force,
and the inward forces due to ram pressure of the rapid inflow and the BH's gravity.
We here assume that (i) the shell is optically thick to the UV
  (ionizing) radiation and absorbs all incident radiation with
  momentum of $L/c$, and that (ii) the entire cloud is effectively
  optically thin to the recombination radiation.  If the recombination
  radiation is efficiently scattered by the neutral shell, that is, if
  condition (ii) is invalid, then the radiation is trapped within the
  shell (i.e. the neutral shells just outside the \hii~region).
  Multiple scattering events in this regime would increase the total
  radiation pressure force to $\simeq \tau_{\rm scat} L/c$, where
  $\tau_{\rm scat}$ is an effective optical depth to scattering.  In
  our case, HI Rayleigh scattering is negligible, but Ly$\alpha$
  scattering would be important because of the high optical depth at
  the line center, $\tau_{\rm Ly\alpha}\sim 10^{10}-10^{12}$.
  However, before radiation pressure by Ly$\alpha$ affects motion of
  the shell, the Ly$\alpha$ photons would be converted to
  $2S\rightarrow 1S$ continuum photons and $\sim 1$ eV photons (H$^-$
  free-bound transition), to which the shell is optically thin.  We
  therefore expect our condition (ii) to hold, with an effective
  scattering opacity $\tau_{\rm scat}$ at most a factor of a few.
  However, future work is needed to investigate the effect of the
  trapping of Ly$\alpha$ radiation, its conversion to lower-energy
  continuum photons, and the escape of these photons from the clouds.

The growth rate of the shell is given by
\begin{equation}
\ddt{M_{\rm sh}}{t}=\dot{M}\left(1+\frac{\dot{R}_{\rm sh}}{|v|}\right),
\end{equation}
and the initial shell mass is given by
\begin{equation}
M_{\rm sh,0}=\int^{R_{\rm sh,0}}_0 4\pi r^2\rho(r) \dd r,
\end{equation}
where the subscript 0 means the initial value.
For simplicity, we consider two extreme cases for the density profile: a constant density profile 
$\rho(r)=\text{const.}$ and the Bondi profile $\rho(r)\propto r^{-3/2}$, with corresponding initial
masses of
\begin{align}
M_{\rm sh,0}=
\begin{cases}
\displaystyle \frac{4}{3}\pi R_{\rm sh,0}^3\rho_\infty & \text{for $\rho(r)=\rho_\infty$} \vspace{2mm} ,\\ 
\displaystyle \frac{8}{3}\pi R_{\rm B}^{3/2}R_{\rm sh,0}^{3/2}\rho_\infty & 
\displaystyle\text{for}~\rho(r)=\rho_\infty\left(\frac{r}{R_{\rm B}}\right)^{-3/2}. \label{eq:dens profile}
\end{cases}
\end{align}
We here set $\dot{M}=\dot{M}_{\rm B}$, the free-fall velocity $|v|=(2GM_{\rm BH}/r)^{1/2}$,
$\dot{R}_{\rm sh,0}=0$,
$M_{\rm BH}=2\times10^4~\msun$, $n_\infty=10^5~{\rm cm^{-3}}$ and $T_\infty=10^4~{\rm K}$.

\begin{figure}
\begin{center}
\includegraphics[width=75mm]
{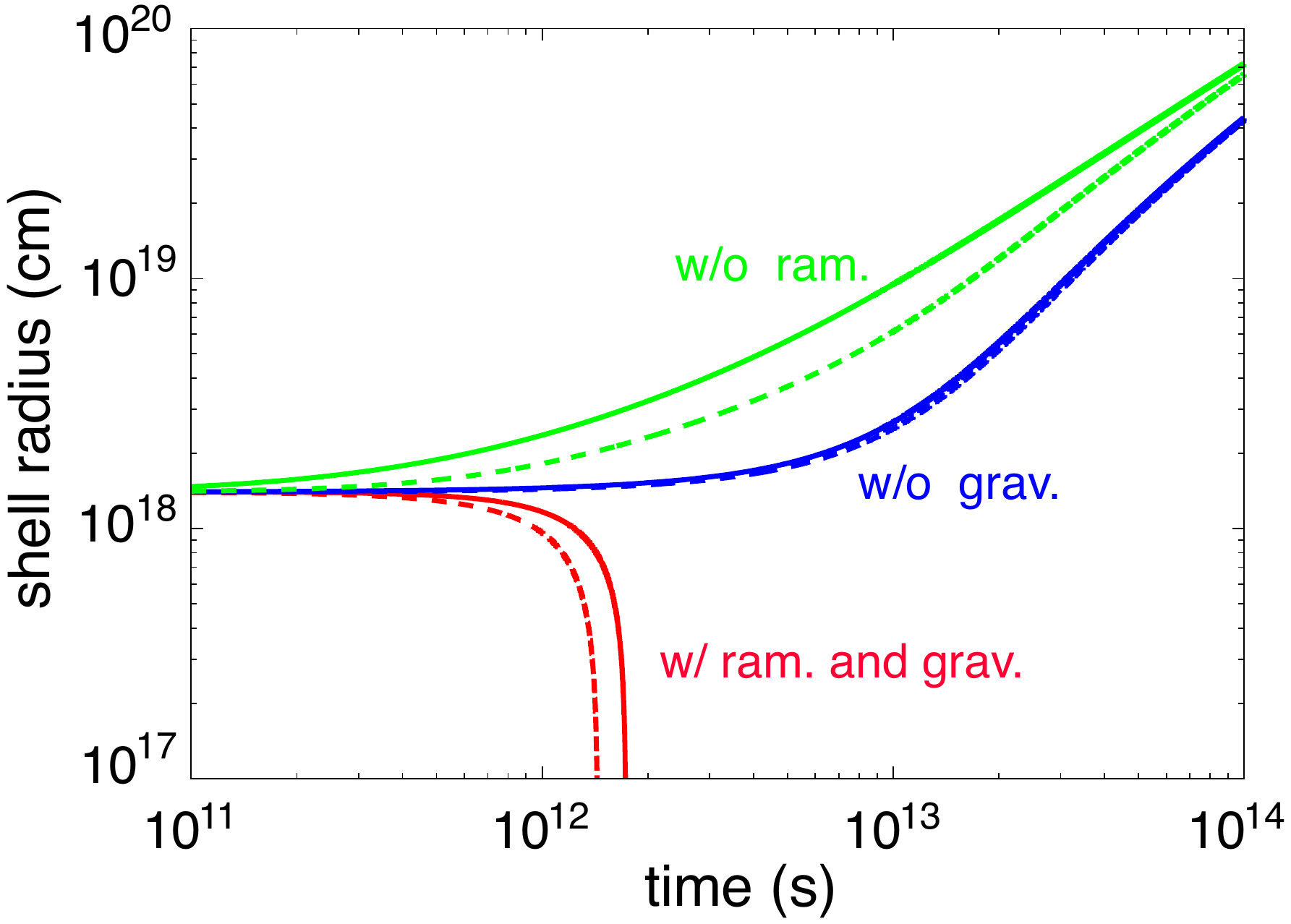}
\caption{Time evolution of a geometrically thin, optically thick
  shell, driven by the radiation force of a central source into a
  rapidly collapsing cloud.  The evolution is from a toy model that
  incorporates the outward radiation force, as well as the
  inward ram pressure and gravitational forces on the shell, initially
  located at $R_{\rm sh,0}=R_\text{\hii}~(\simeq 1.4\times10^{18}~{\rm
    cm})$.  Each curve corresponds to the case with both ram
  pressure and gravity (red), and with either gravity (blue) or ram
  pressure (green) artificially turned off.  In each case, the initial density
  profile was assumed to be either constant (solid) or to follow the Bondi
  profile (dashed; Eq.~\ref{eq:dens profile}).  We set
  $\fedd=1$, $\dot{M}=\dot{M}_{\rm B}$, $M_{\rm
    BH}=2\times10^4~\msun$, $n_\infty=10^5~{\rm cm^{-3}}$ and
  $T_\infty=10^4~{\rm K}$.  This shell corresponds to that shown in
  phase (i) in Fig.~\ref{fig:hist1}(a).}
  \label{fig:shell1}
\end{center}
\end{figure}

First, we consider time evolution of a dense shell which initially stalls at
$R_{\rm sh,0}=R_\text{\hii}~(\simeq 1.4\times10^{18}~{\rm cm}$)
before the transition to hyper-Eddington accretion occurs, when $L\simeq \ledd$ ($\fedd \simeq 1$).
This shell corresponds to that shown in Fig.~\ref{fig:hist1}(a) (phase 1).
Fig.~\ref{fig:shell1} shows three cases, in which the ram pressure of the inflowing gas and the BH's gravity on 
the accumulated mass of the shell are both included (red), and in which either the gravity (blue) or the ram pressure
(green) are artificially turned off.
Solid (dashed) curves correspond to constant (Bondi) initial density profiles.
As this figure shows, when both ram pressure and gravity are included, the shell radius contracts.
On the other hand, when either of the inward forces are turned off the shell continues to expand, and
never accretes onto the center.
Note that the expansion velocity of the shell is slower for the cases with heavier masses (dashed),
but the choice of the initial shell mass is not important.
Overall, we infer that it is the combination of the ram pressure and gravity
that overwhelms radiation force and yields hyper-Eddington accretion.
The role of ram pressure is found to be somewhat more important (the shell expands faster without ram pressure [green] than without gravity [blue]).

\begin{figure}
\begin{center}
\includegraphics[width=73mm]
{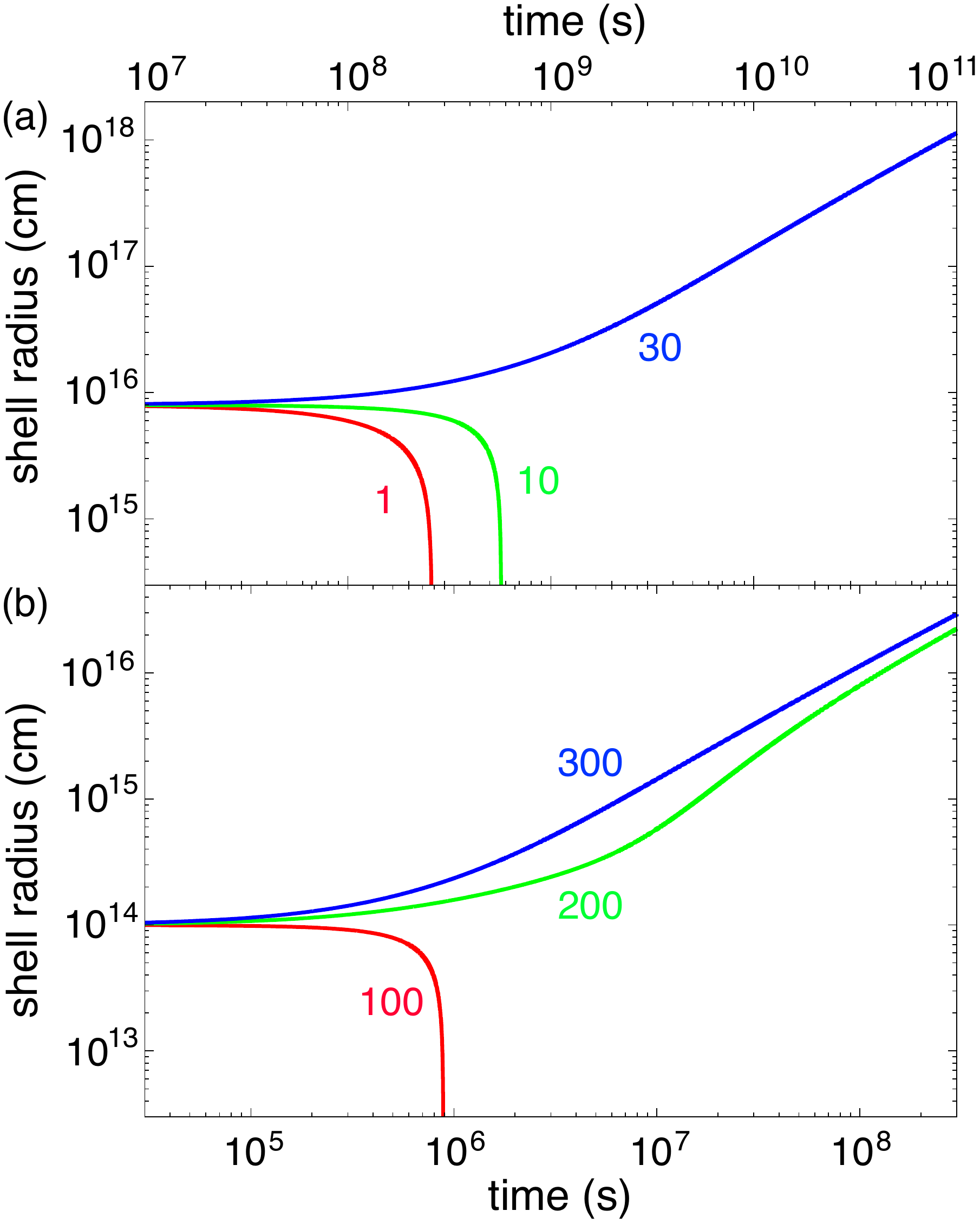}
\caption{
Same as Fig.~\ref{fig:shell1} but for different values of $R_{\rm sh,0}$ and $\fedd$. 
 The initial shell radii and $\fedd$ are set to (a) $R_{\rm sh,0}=r_{\rm min}~(=8\times10^{15}~{\rm cm})$ and 
 $\fedd=1$ (red), $10$ (green) and $30$ (blue) or to
 (b) $R_{\rm sh,0}=R_{\rm ph}~(\simeq 10^{14}~{\rm cm})$ and $\fedd=100$ (red), $200$ (green) and $300$ (blue).
 The initial shell mass is computed for a constant density profile, and the effects of ram pressure and gravity are both included.
}
  \label{fig:shell2}
\end{center}
\end{figure}

Next, Fig.~\ref{fig:shell2}(a) shows the time evolution of a shell initially located at
$R_{\rm sh,0}=r_{\rm min}(=8\times10^{15}~{\rm cm})$, for $\fedd=1$ (red), $10$ (green) and $30$ (blue).
These correspond to the cases after hyper-Eddington accretion is realized in the simulations.
We here estimate the initial shell mass assuming a constant density profile, and
the effects of ram pressure and gravity are both included.
Fig.~\ref{fig:shell2}(a) clearly shows that for $\fedd \la10$, the shell contracts within $20$ yr,
resulting in hyper-Eddington accretion.
This result is in excellent agreement with our simulations and analytical arguments in \S\ref{subsec:analytical1}.
Fig.~\ref{fig:shell2}(b) also shows the case for the initial shell radius of
$R_{\rm sh,0}=R_{\rm ph}(\simeq 10^{14}~{\rm cm})$,
for $\fedd=100$ (red), $200$ (green) and $300$ (blue).
The shell can contract for $\fedd\lesssim100$, which is again in agreement with 
the results shown in Figs.~\ref{fig:condition1}~and~\ref{fig:condition2}.

\section{Summary and discussions}
\label{sec:summary}

We have performed one-dimensional radiation hydrodynamical simulations
to solve spherically symmetric accretion flows onto massive BHs with a
very high rate.  Our setup extends simulations in our earlier work
\citetalias{Inayoshi2015a}, by allowing the central luminosity to
exceed the Eddington luminosity ($1\la L/\ledd \la 100$).  This is
motivated by the possibility of gas accreting with finite angular
momentum, and forming a bright nuclear disc, fed at rates well in
excess of the Eddington rate \citep[e.g.][]{Ohsuga2005,Sadowski2014,Jiang2014,Mckinney2014}.

We find that a transition to hyper-Eddington accretion phase occurs when
\begin{equation}
  \dot{M}\ga 5000~\dot{M}_{\rm Edd}=5000~L_{\rm Edd}/c^2,
  \label{eq:oldcond}
\end{equation}
(see also equation \ref{eq:condition}). 
This condition remains identical to that found by \citetalias{Inayoshi2015a} who
assume that photon trapping effectively limits the emerging luminosity to $\la L_{\rm Edd}$.
However, we identify a new condition, which determines whether the hyper-Eddington accretion is steady,
or strongly episodic.  We find that a steady state is maintained 
as long as the radiation luminosity from the central source is below a critical value,
\begin{equation}
\frac{L}{\ledd}\la 11~\left(\frac{M_{\rm BH}}{10^4~\msun}\right)^{3/2}
\left(\frac{n_{\infty}}{10^5~\cc}\right).
\end{equation}
This corresponds to the requirement that the ram pressure of the collapsing medium and the BH gravity on the accumulated mass of the shell
at the edge of the \hii~region dominate over the radiation force, i.e. $\dot{M}|v|\gtrsim L/c$
(see equation \ref{eq:ram}).
If the luminosity exceeds this critical value, then a steady
hyper-Eddington phase can not exist, and the accretion instead becomes
episodic. The time-averaged rate still matches the unimpeded Bondi
rate, $\dot{M}_{\rm B}$, provided that the condition in
Eq.~\ref{eq:oldcond} is satisfied.
We summarize the three different types of accretion flows, determined by the above two conditions, in Figs.~\ref{fig:condition1} and \ref{fig:condition2}.
In this paper, we also offered a physical understanding of our simulation results: we showed that the latter can be recovered in a toy model of an optically thick spherical shell, driven by radiation into a collapsing cloud.

Throughout this paper, we have assumed a single power-law radiation spectrum with an index of $\alpha=-1.5$ 
over a frequency range of $13.6~{\rm eV}\leq h\nu \leq 30~{\rm keV}$ (see equation \ref{eq:power}).
In this case, all of the radiation can contribute to ionization of neutral gas.
However, a realistic spectrum of an accretion disk around a BH would 
allows lower energy photons with $h\nu < 13.6~{\rm eV}$ which 
can escape without ionizing the accreting gas
(e.g., \citealt{Tanaka2010}).
Thus, assuming a realistic spectrum, but with a fixed total luminosity,
the transition to hyper-Eddington accretion is more likely to occur,
compared to our case.

In our simulations, the photosphere and trapping radius, located at small radii, are not resolved directly.
When the accreting gas has a finite angular momentum, a compact accretion disk should form around the central BH.
Then, anisotropic radiation and/or outflows/jets from the center would break the spherical symmetry of the inflow,
at least in the inner regions.  A fully self-consistent treatment of such an accretion flow, which has an anisotropic,
bright source with $L>L_{\rm Edd}$ at the central region, embedded in a quasi-spherical inflow at large radii,
will require multi-dimensional radiation hydrodynamical simulations.

\section{Acknowledgements}

Yuya Sakurai thanks the Columbia University Astronomy Department for
its hospitality during an extended visit, during which this work was
completed.  We thank Mark Dijkstra, Jeremiah P. Ostriker and Naoki Yoshida for fruitful discussions.  
This work is partially supported by Advanced Leading Graduate Course for
Photon Science (YS), by Grant-in-Aid for JSPS Fellows (15H00776: YS),
by the Simons Foundation through the Simons Society of Fellows (KI),
and by NASA grants NNX11AE05G and NNX15AB19G (ZH).  Numerical
computations were carried out on a PC cluster at the Center for
Computational Astrophysics of the National Astronomical Observatory of
Japan.

\small{\bibliography{ms}}

\end{document}